\documentclass{ckm}                 

\usepackage{txfonts}            
\usepackage{epsfig}

\newcommand{\BaBar}{{\sc BaBar}}
\newcommand{\Vub}{\mbox{$|V_{ub}|$}}
\newcommand{\Vcb}{\mbox{$|V_{cb}|$}}
\newcommand{\sintwob}{\mbox{$\sin 2 \beta$}}
\newcommand{\deltamd}{\mbox{$\Delta m_d$}}
\newcommand{\bbbar}{\mbox{$B^0 \bar B^0$}}

\newcommand{\BXulv}{\mbox{$B \to X_u \ell \nu$}}
\newcommand{\BXclv}{\mbox{$B \to X_c \ell \nu$}}
\newcommand{\BXlv}{\mbox{$B \to X \ell \nu$}}
\newcommand{\BDstlv}{\mbox{$B^0 \to D^{*-} \ell^+ \nu$}}
\newcommand{\Dsmunu}{\mbox{$D^+_s \to \mu^+ \nu$}}
\newcommand{\pl}{\mbox{$p_{\ell}$}}
\newcommand{\mes}{\mbox{$m_{\rm ES}$}}
\newcommand{\mx}{\mbox{$m_X$}}
\newcommand{\mxs}{\mbox{$m_X^2$}}
\newcommand{\mb}{\mbox{$m_b$}}
\newcommand{\lambdaone}{\mbox{$\lambda_1$}}
\newcommand{\Lambdabar}{\mbox{${\bar \Lambda}$}}
\newcommand{\BXsg}{\mbox{$B \to X_s \gamma$}}
\newcommand{\bsg}{\mbox{$b \to s \gamma$}}

\confname{Workshop on the CKM Unitarity Triangle, IPPP Durham, April
 2003}

\title{Experimental Review of Inclusive \Vub\ Measurements}

\author{F Muheim\thanks{Email: F.Muheim@ed.ac.uk}}

\address{School of Physics, University of Edinburgh}

\begin{document}

\begin{abstract} We review the status
of inclusive \Vub\ measurements where many new results have become
available recently.

\end{abstract}

\maketitle


\section{Introduction}

Precise measurements of the CKM matrix elements 
\Vub\ and \Vcb\ provide stringent consistency tests of the 
unitarity triangle. The data on the ratio \Vub/\Vcb\  
form an annulus in the unitarity plane. This  can be tested
against the apex of the triangle, determined by the measurements
of the CP violating phase \sintwob\ and the \bbbar\ mixing parameter
\deltamd. The \BaBar\ and Belle experiments have recently measured
 \sintwob\ with an error of  0.055~\cite{ref:gerhard}
and this precision is expected to improve to 0.02 in a few years time.
The precision on \Vub\ will soon limit unitarity triangle tests.
Hence we aim to measure \Vub\ to 10\% or better 
in the near future.

\section{Status and Methods}

Ten Years ago, CLEO and Argus had established that \Vub\ was
non-zero~\cite{ref:cleo93cleo89argus89} by observing
an excess of leptons at the endpoint of the momentum spectrum
(2.3 - 2.6 GeV/c).
However, the accuracy of this result, 
$ \Vub/\Vcb = 0.08 \pm 0.02$, was theoretically limited
by the extrapolation into the full phase which were based on (exclusive) 
models. 
Since then huge progress has been made by both, theory and experiment.
As reported at this workshop~\cite{ref:luke},
Heavy quark effective theory (HQET)  and Operator production expansion (OPE)
allow a calculation of the inclusive  charmless semileptonic 
decay rate of $B$ mesons, \BXulv. The distributions for several kinematic
variables, e.g. the lepton momentum, \pl, 
have also been calculated
up to order $\alpha_S$~\cite{ref:neubert-defazio}.
The theoretical uncertainties in the determination of \Vub\ 
on the $b$-quark mass, \mb,  and the average kinetic energy
of the $b$-quark within the $B$ meson, \lambdaone,
are now quantifiable. 

The experimental challenge is to suppress the background
from \BXclv\ transitions. These have a rate that is
higher by two orders of magnitude.
Only $\sim 10\%$ of \BXulv\ decays have a lepton momentum
larger than 2.3 GeV/c and are above 
the endpoint of the \BXclv\ background.
With the large statistics of $B$ mesons available at 
the $B$ factories, additional methods with lower efficiency but higher
purity are now accessible.

An additional kinematic variable has been already employed  by
the LEP experiments, namely the
mass \mx\ of the hadronic system $X_u$ recoiling against the lepton pair
in the rest frame of the $B$ meson.
About 70\% of the \BXulv\ decays have  $\mx < m_D$ 
where the mass of the
$D$ meson, $m_D$, is the minimum mass \mx\ for \BXclv\ decays.
In $\sim$ 25\% of \BXulv\ decays
the invariant mass of the lepton-neutrino pair, $q^2$, 
exceeds the maximum possible value of background \BXclv\ decays. 
The energy of the hadronic recoil system could also be used to
separate signal from background.
The distributions  of these kinematic 
variables have different dependencies on \mb\ and \lambdaone.
Thus it is important to measure all kinematic distributions.

Here, we  give an overview of six new results on inclusive 
\Vub\ measurements  which have been made public
by the CLEO, \BaBar\ and Belle collaborations during the last year.
For details on these measurements the reader is referred to
other presentations at this workshop~\cite{ref:sarti},\cite{ref:kakuno}.
Measured inclusive branching ratios \BXulv\ are converted into
\Vub\ measurements using the relation
$$ \Vub = 0.00445 
\sqrt{ \frac{{\cal B}(BXulv)}{0.002} \frac{1.55 \rm ps}{\tau_B}}
\left( 1 \pm 0.020 \pm 0.052 \right) $$
or equivalent~\cite{ref:uraltsev},\cite{ref:hoangetal},\cite{ref:pdg2002}.
The first error arises from the perturbative scale dependence and
the second error is due to the  $b$ quark mass for which
an uncertainty of  90~MeV has been assigned. 
In what follows,
we will refer to these two uncertainties as theoretical error.

\section{Lepton Endpoint Spectrum}

A new measurement on the excess of leptons in the endpoint region
has been published by CLEO~\cite{ref:cleo-endpoint}.
The full data set of 9.13~fb$^{-1}$ at the $\Upsilon(\mathrm 4S)$ resonance
and 4.35~fb$^{-1}$ just below has been used.
They have carefully designed  the suppression 
of the background from non-$B$ decays (continuum)
in order to minimise model dependence on the selection cuts.
\begin{figure}[htb]
\hbox to\hsize{\hss
\includegraphics[width=\hsize]{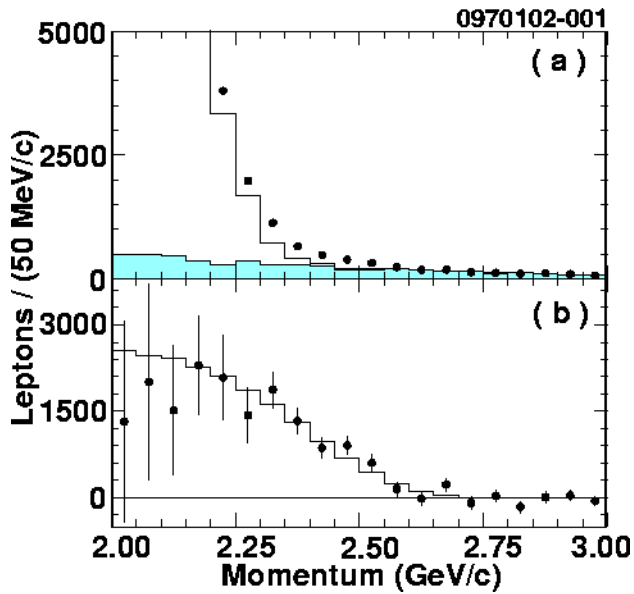}
\hss}
\caption{(a) CLEO lepton spectra for on- (points) and scaled 
off-resonance data (shaded histogram). The unshaded 
histogram is the sum of the scaled off-resonance
and $B$-decay background. (b) Background subtracted and efficiency
corrected lepton spectrum for \BXulv\ (points) and simulation (histogram). 
}
\label{fig:cleo-endpoint}
\end{figure}
In Fig.~\ref{fig:cleo-endpoint}  
we show the resulting lepton momentum spectrum.
A signal interval of 2.2 - 2.6 GeV/c was chosen as this
gives the smallest total error.
A clear excess of $1901 \pm 122 \pm 256$ events, attributed to
\BXulv, is visible in this region. 
A partial branching ratio of
%
$$\Delta {\cal B}(\BXulv )  = 
\left(0.230 \pm 0.015 \pm 0.035 \right) \times 10^{-3}
$$
%
%
%
is measured.
The systematic error is dominated by the uncertainty in
the \BXclv\ background subtraction.
CLEO uses their own measurement of the \BXsg\ energy spectrum
~\cite{ref:cleo-moments} 
to determine the universal shape function
of the heavy-light quark pair 
system~\cite{ref:luke},\cite{ref:neubert-defazio}.
This in turn allows to calculate the fraction of \BXulv\ signal 
in the above interval, $f_u = 0.130 \pm 0.024 \pm 0.015$,
and to extract a \BXulv\ branching ratio.
The branching ratio measurement is then 
converted into a \Vub\ result and CLEO obtains
$$ \Vub = \left( 4.08 \pm 0.34 \pm 0.44  
\pm 0.16   \pm 0.24  \right) \times 10^{-3}
$$
where the errors are separated into experimental, 
signal fraction ($f_u$),
theoretical and shape function components, respectively.

Using  the lepton endpoint method, 
The \BaBar\ experiment has presented a preliminary measurement 
based on 20.6~fb$^{-1}$ of data at the $\Upsilon(\mathrm 4S)$ 
resonance and
2.6~fb$^{-1}$ below~\cite{ref:babar-endpoint}. 
An excess of $1696 \pm 133 \pm 153$ events is observed in
the lepton momentum range of 2.3 - 2.6 GeV/c.
\BaBar\ measures a partial branching ratio of
%
$$\Delta {\cal B}(\BXulv )  = 
\left(0.152 \pm 0.014 \pm 0.014 \right) \times 10^{-3}
$$
%
%
%
where the experimental systematic error is dominated by the uncertainty in
the continuum background subtraction.
The CLEO measurement of $f_u = 0.074 \pm 0.014 \pm 0.009$
for this momentum interval is used to extrapolate to 
the full phase space. Due to the larger extrapolation
the uncertainty is comparable to the experimental systematic error. 
From the \BXulv\ branching ratio 
\BaBar\ measures a preliminary \Vub\ result of
$$ \Vub = \left( 4.43 \pm 0.39 \pm 0.50  
\pm 0.25   \pm 0.35  \right) \times 10^{-3}
$$
where the errors are separated into experimental, signal fraction,
theoretical and shape function components, respectively.

\section{Hadronic Recoil Mass}

\begin{figure}[htb]
\hbox to\hsize{\hss
\includegraphics[width=\hsize]{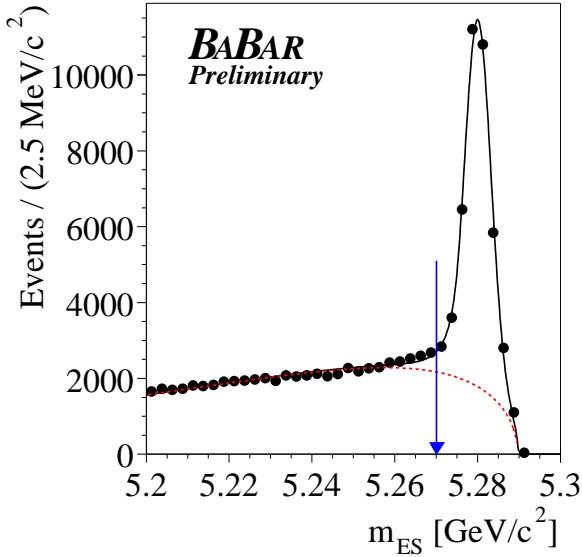}
\hss}
\caption{\BaBar\ \mes\ distribution for hadronic tagging $B$ mesons 
in the lepton sample. }
\label{fig:babar-mes}
\end{figure}

A  measurement of \Vub\ by reconstructing
the mass \mx\ of the hadronic system 
recoiling against the lepton pair
in the \BXlv\ decay has been carried out by 
\BaBar\ \cite{ref:sarti},\cite{ref:babar-mx}.
A data sample of about $82 \; \rm fb{-1}$ is used.
The large statistics of $B$ meson samples produced at the B-factories
permits the use of new methods which have low efficiency, but 
allow to exploit the separation power of \mx.
The experimental task is to separate the 
spatially overlapping decay products of the two $B$ mesons produced at the 
$\Upsilon(\mathrm 4S)$.
The principle is to fully reconstruct one $B$ meson in many
hadronic modes:
$B \to D^{(*)} \pi, \; D^{(*)} \pi \pi^0, \; D^{(*)} 3\pi, \; ...$. 
The hadronic $B$ meson tags the $b$-quark flavour, and more importantly
removes the background from one of the $B$ mesons in the event.
The method yields about 4000 $B$ mesons per $\rm fb^{-1}$.
All the remaining particles that are detected in a tagged event must
originate from the other $B$ meson.
Events are selected if a lepton with a 
momentum $\pl > 1 \; \rm GeV/c$ in the $B$ rest frame is identified
among the particles recoiling 
against the tagging $B$ meson.
The energy substituted mass spectrum, \mes, 
of the tagging $B$ candidates 
is shown in Fig.~\ref{fig:babar-mes}. The signal to background ratio
is large and a fit is used to remove 
the continuum background.
Additional criteria, e.g. charge correlation
and a missing mass consistent with a neutrino are applied
to increase the purity of this sample.
The hadronic recoil mass, \mx, is reconstructed;
a constraint fit imposing zero missing mass 
is employed to improve the \mx\ resolution.
The \BXulv\ signal region 
is also enriched by vetoing semileptonic $B$ candidates
containing $K^{\pm}$ or $K^0_S$ decay products. The depleted region
is used as a control sample.
\begin{figure}[htb]
\hbox to\hsize{\hss
\mbox{\epsfig{figure=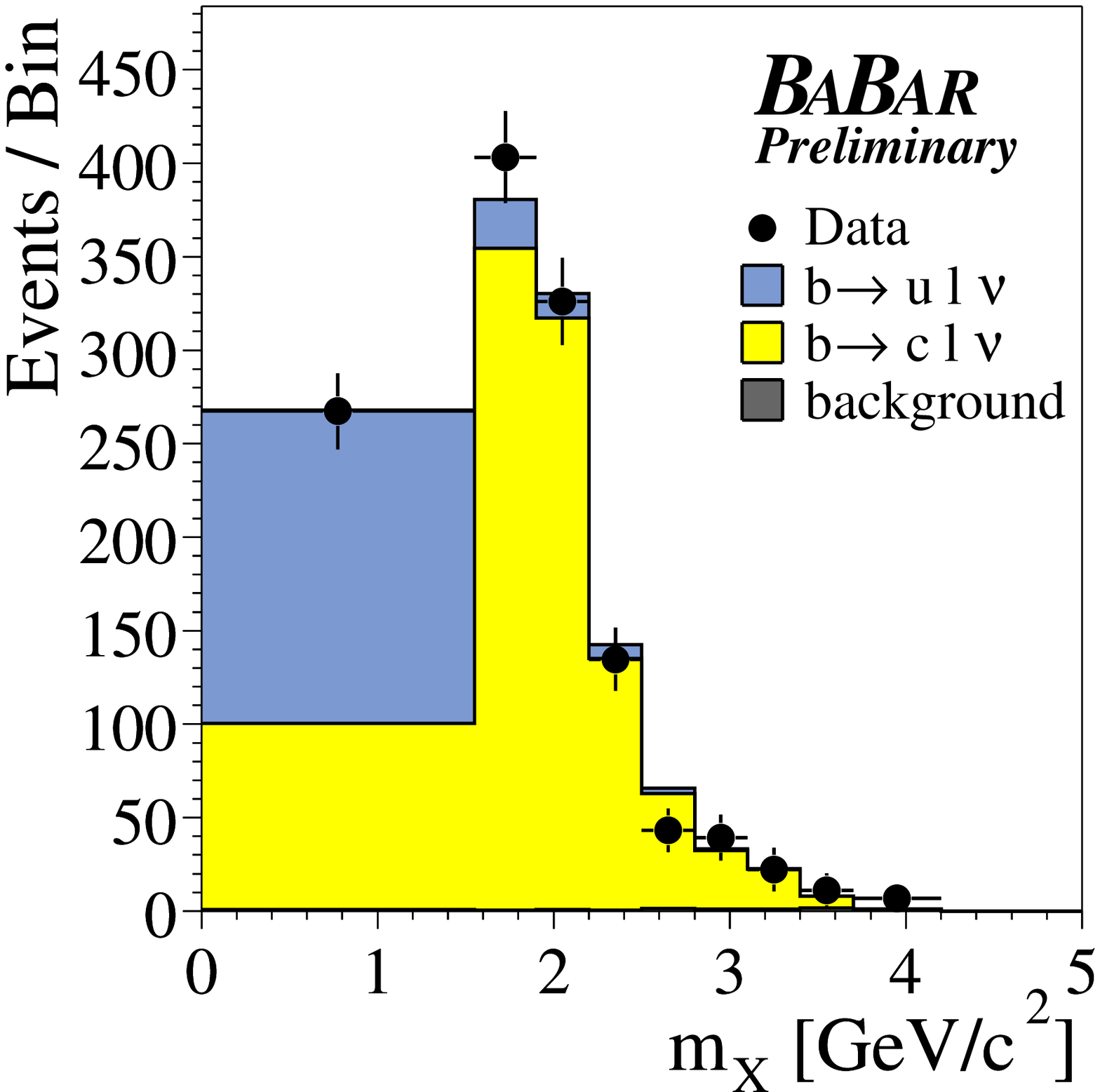,width=0.49\linewidth}	
	\epsfig{figure=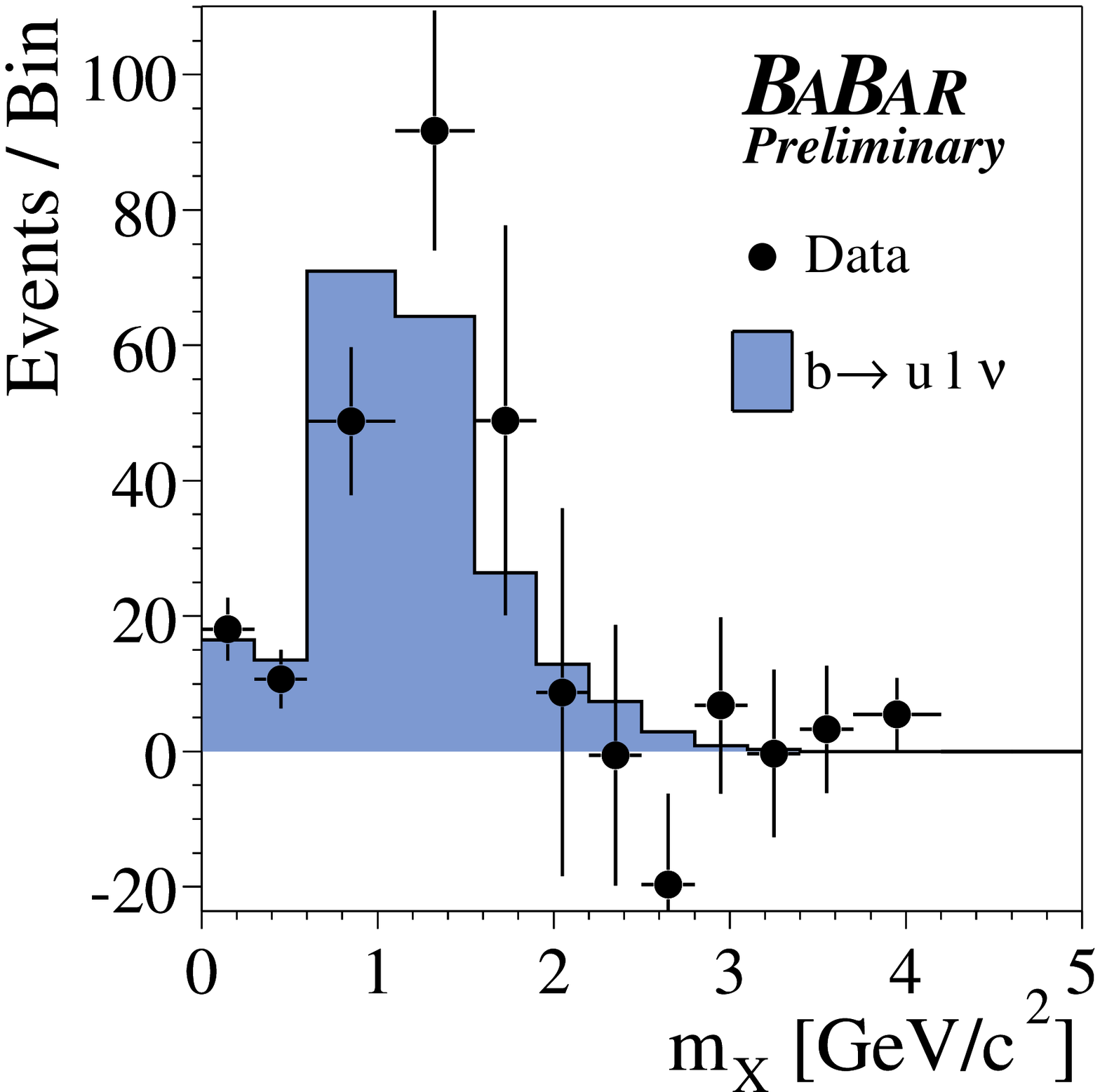,width=0.49\linewidth}}
\hss}
\caption{(a) \BaBar\ \mx\ spectrum (dots) and simulation (histograms).
(b) Background subtracted data (dots) and \BXulv\ simulation overlaid.  }
\label{fig:babar-mx}
\end{figure}
In Fig.~\ref{fig:babar-mx}a) we plot the measured 
\mx\ spectrum. A binned fit is overlaid. 
The signal region is $\mx < 1.55 \; \rm GeV$ which gives
the smallest total error. It contains one bin to minimise the model dependence.
 An excess of $167 \pm 21$ events is observed and attributed
to \BXulv. The signal to background ratio is about 2:1.
The background subtracted \mx\ spectrum with finer bins in
the signal region is shown 
in Fig.~\ref{fig:babar-mx}b).
To reduce systematic errors, 
the ratio $R_u = {\cal B}(\BXulv)/{\cal B}(\BXlv)$ is measured.
\BaBar\ obtains a preliminary result of
$$ \frac{{\cal B}(\BXulv)}{{\cal B}(\BXlv)}
        = \left( 1.97 \pm0.27 \pm 0.23  
\pm 0.34\right) \times 10^{-2}
$$
where the errors are statistical, systematic and 
extrapolation into the region above the \mx\ cut.
Using this branching ratio measurement \BaBar\ presents the following
preliminary \Vub\ result
$$ \Vub = \left( 4.52 \pm 0.31 \pm 0.27 
\pm 0.30   \pm 0.26  \right) \times 10^{-3}
$$
where the errors are statistical, systematic, 
\mx\ extrapolation and theoretical.
The extrapolation error still dominates, but improvements in the knowledge
of \Lambdabar\ and \lambdaone\ are expected 
from measurements in the next few years.
The potential of this method is its small systematic error, thus
more statistics will allow to improve the precision on \Vub.

\begin{figure}[htb]
\hbox to\hsize{\hss
\includegraphics[width=\hsize]{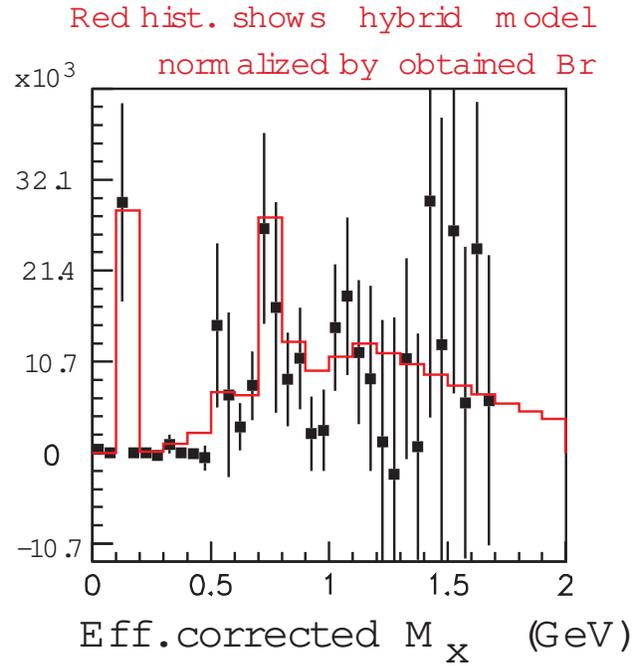}
\hss}
\caption{Background subtracted \mx\ spectrum (squares) from Belle. The 
\BXulv\ simulation (histograms) is overlaid.}
\label{fig:belle-mx}
\end{figure}
Employing a method similar to \BaBar,
the Belle experiment has presented a preliminary \Vub\ 
result~\cite{ref:kakuno},\cite{ref:belle-mx}. 
Fully reconstructed \BDstlv\ decays are used as tagging $B$ mesons.
Hence, in selected events, both mesons decay semileptonically.
The tag removes the decay products of 
one of the $B$ mesons and substantially suppresses the continuum 
background from non-$B$ decays. 
Since there are two neutrinos in the event, the $B$ momentum 
cannot be determined  from the tagging $B$ meson.
The direction of the two $B$ momenta lay on two cones which can be 
reconstructed up to an azimuthal angle. 
After mirroring one of the cones on the origin, 
the $B$ direction must then lay on the cross line of the 
two cones and can be reconstructed.
As with hadronic tags,
the hadronic recoil mass \mx\ 
is used to separate
\BXulv\ decays from \BXlv\ decays.
In the signal region, defined as $\mx < 1.5\; \rm GeV$,
Belle observes an excess of
$172 \pm 28$ leptons with momenta $\pl > 1 \; \rm GeV/c$
as is shown in Fig.~\ref{fig:belle-mx}. 
This excess correspond to a  branching ratio
$$ {\cal B}(\BXulv)
        = \left( 2.62  \pm 0.63 \pm 0.23  
\pm 0.41\right) \times 10^{-3}
$$
where the errors are statistical, systematic and 
extrapolation into the region above the \mx\ cut.
Belle obtains a preliminary result for \Vub\ of 
$$ \Vub = \left( 5.00 \pm 0.60 \pm 0.23
\pm 39 \pm 0.36   \right) \times 10^{-3}
$$
where the errors are separated into statistical, experimental,
\mx\ extrapolation and theoretical components, respectively.

\section{Neutrino Reconstruction}

The neutrino reconstruction
technique relies on exploiting the hermeticity of the detector
to infer the neutrino momentum from the missing momentum.
The missing energy is also measured and allows to require that 
the missing mass-squared of the event be 
consistent with zero.
For experiments at the 
$\Upsilon(\mathrm 4S)$ resonance, this method has been pioneered
for exclusive \Vub\  and \Dsmunu\ measurements
~\cite{ref:gibbons},\cite{ref:dsmunu}.

\begin{figure}[htb]
\hbox to\hsize{\hss
\includegraphics[width=\hsize]{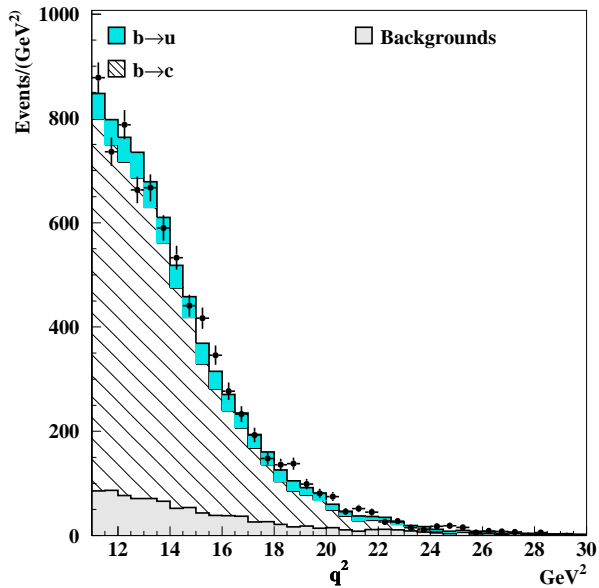}
\hss}
\caption{Fit projection for the $q^2$ spectrum from 
CLEO (points).
The \BXulv\ signal, \BXclv\ and other background simulations
 are shown as histograms.}
\label{fig:cleo-neutrino}
\end{figure}
Using neutrino reconstruction,
the CLEO experiment has presented a preliminary result
on \Vub\ using events with inclusive charged leptons~\cite{ref:cleo-neutrino}.
The measurement of the neutrino momentum allows the 
use of kinematic variables in addition to the 
charged lepton energy, $E_{\ell}$.
The hadronic recoil mass squared, \mxs,
the invariant mass of the lepton pair, $q^2$,
and $\cos \theta_{W\ell}$, the helicity angle of the virtual $W$
are used here.
Selected events are required to have a lepton with momentum $\pl >  1$~GeV/c.
A simultaneous fit to the quantities 
$q^2$, \mxs, and $E_{\ell}$ is performed.
The \BXulv\ signal region is restricted to the range
$q^2 > 10 \; \rm GeV^2$ and $\mxs < 2.25 \; \rm GeV^2$.
In Fig.~\ref{fig:cleo-neutrino} we show the $q^2$ projection 
of the fit with the \mx\ cut applied for a data sample of 9.4~fb$^{-1}$.
From this fit CLEO extracts  a preliminary result of
$$ \Vub = \left( 4.05 \pm 0.18 \pm 0.58  
\pm 0.33   \pm 0.56  \right) \times 10^{-3}
$$
where the errors are statistical, experimental,
\mx\ and $q^2$ extrapolation and  theoretical.

\begin{figure}[hqtb]
\hbox to\hsize{\hss
\includegraphics[width=\hsize]{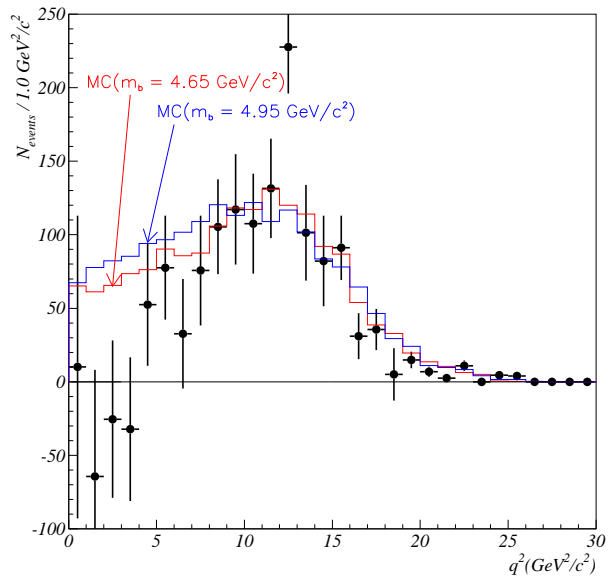}
\hss}
\caption{Background subtracted $q^2$  spectrum (points) from Belle. The 
\BXulv\ simulation (histograms) is overlaid.}

\label{fig:belle-neutrino}
\end{figure}
The Belle experiment has also presented a preliminary 
\Vub\ result using events with charged leptons and 
neutrino reconstruction~\cite{ref:kakuno},\cite{ref:belle-mx}.
In addition, the particles in each event are separated into a 
tagging and a signal $B$ meson
by employing a combinatorial annealing~\cite{ref:annealing}.
The annealing procedure minimizes the likelihood ratio of random
over the correct assignments of particles to the two $B$ mesons
in the events. The efficiency of this method is about 0.3\%.
A control sample of \BDstlv\ decays is used
to check the annealing and to correct for efficiency differences between
data and simulation. 
After annealing, all kinematic variables of interest
are available, 
for example $q^2$ and the hadronic recoil mass \mx\ in the $B$ mesons
decaying semileptonically.
The signal is extracted in the region of
$q^2 > 7 \; \rm GeV^2$ and $\mx < 1.5 $ GeV.
Belle measures 
an excess of 1148 events over background is measured with a signal to 
background ratio of 0.27.
In Fig.~\ref{fig:belle-neutrino} we plot the background
subtracted $q^2$ spectrum 
with  the \mx\ cut applied. Also shown are \BXulv\ simulations
for two values of \mb.
A preliminary branching ratio of 
$$ {\cal B}(\BXulv)
        = \left( 1.64  \pm 0.14 \pm 0.36  
\pm 0.36\right) \times 10^{-3}
$$
is measured.
The errors are statistical, systematic and 
extrapolation into the region above the \mx\ and $q^2$ cut.
From this measurement,
Belle obtains a preliminary \Vub\ result of 
$$ \Vub = \left( 3.96 \pm 0.17 \pm 0.44  
\pm 0.43   \pm 0.29  \right) \times 10^{-3}
$$
where the errors are separated into statistical, experimental,
\mx\ and $q^2$ extrapolation and theoretical components, respectively.

\section{Summary and Conclusions}

The B-factories experiments, \BaBar\ and Belle, and CLEO have
recently produced many new results for the CKM matrix element \Vub.
Different  kinematic variables have been used in the results which 
are based on 
measurements of the inclusive \BXulv\ decay rate.
\begin{figure}[hqtb]
\hbox to\hsize{\hss
\includegraphics[width=\hsize]{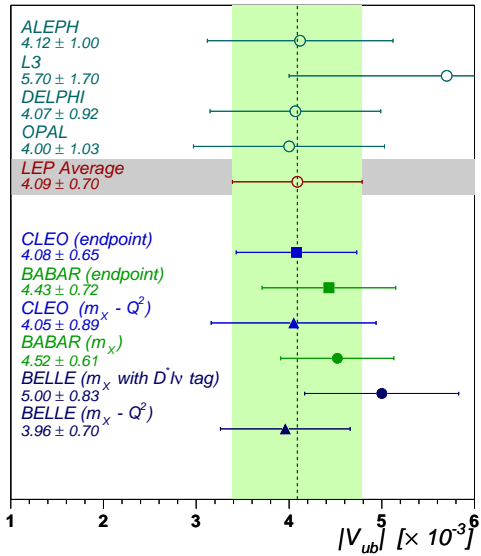}
\hss}
\caption{Compilation of inclusive \Vub\ results by the 
Heavy Flavour Averaging Group for the winter conferences 2003.}
\label{fig:hfag}
\end{figure}
In Fig.~\ref{fig:hfag}, we show a compilation of all 
\Vub\ results, carried out by
 the Heavy Flavour Averaging Group (HFAG) for
the winter conferences 2003~\cite{ref:hfag}.
A procedure to average these results is being 
developed. We have made a preliminary average of all 
inclusive \Vub\ measurements made by the \BaBar, Belle and CLEO
experiments and obtain
$$ \Vub = \left( 4.32 \pm 0.57 \right) \times 10^{-3} \; .
$$
Statistical and detector errors are treated uncorrelated whereas 
the extrapolation and theoretical errors are correlated and 
rescaled.
A arbitrary 5\% error is assigned to the validity of the shape
function being equal for \bsg\ and \BXulv. 
Hence we consider this a conservative average.

In this review, we have shown 
that substantial experimental 
and theoretical progress has been made over the last decade
for measuring  \Vub, a crucial parameter
in the unitarity triangle.
The B-factories allow the use of new experimental methods.
The single most precise measurement is
\BaBar\ result using the hadronic recoil mass \mx\ spectrum,
The precision of this result is 14\% which
is better than the average of the LEP results~\cite{ref:LEP}.
Excellent neutrino reconstruction is crucial for measuring
$q^2$ which has lower sensitivity to \mb\ than \mx. 
The theoretical precision will improve as more precise
measurements of the moments of the shape function moments, \Lambdabar\ and \lambdaone, will  become available.
The aim is to reduce the error on \Vub\ to substantially better than 10\%
over the next five years.

\end{document}